# On subwavelength imaging of Maxwell's fish eye lens

Fei Sun[1] and Sailing He[1, 2]

1 Centre for Optical and Electromagnetic Research, Zhejiang University
2 Department of Electromagnetic Engineering, School of Electrical Engineering,
RoyalInstitute of Technology, S-100 44 Stockholm, Sweden

**Abstract**

Both explicit analysis and FEM numerical simulation are used to analyze the field distribution of a line current in the so-called Maxwell's fish eye lens, which has been claimed recently to be able to achieve perfect imaging [4]. We show that such a Maxwell's fish eye lens cannot give perfect imaging due to the fact that high order modes of the object field can hardly reach the image point in the Maxwell's fish eye. If only zero order mode is excited, a subwavelength image can be achieved, however, its spot-size is larger than the spot size of the source field. The image resolution is determined by the field spot size of the image corresponding to the zeroth order component of the object field. Our explicit analysis consists very well with the FEM results for a modified fish eye bounded with perfectly electrical conductor (PEC). Explicit condition is given for achieving a subwavelength image. When this condition is not satisfied, a single line current source may give multiple image spots.

## 1. Introduction

Maxwell's fish eye was proposed by Maxwell in1854 [1]. Maxwell's fish eye gives a good image with equal light paths from the view point of geometry optics [1-3]. Recently, Leonhardt claimed that Maxwell's fish eye can give perfect imaging in wave optics and he modified the original fish eye lens, which is infinitely large, so that the device becomes finite (bounded with a PEC boundary) [4, 5]. Leonhardt gave an explicit solution with very small spot sizes of the object and image fields for a modified fish eye with a line current source in the object point and a line current drain at the image point [4]. However, this configuration is not practical for imaging. For example, we don't know beforehand the distribution of fluorescent points in bio imaging, and thus we cannot determine where to put the active drains in order to achieve an image of excellent resolution. Apparently this is not a conventional concept for imaging. In a conventional image, we consider a very sharp field distribution (produced by some kind of source) and see if a lens can give a very sharp field distribution at another space point (without any active drain). In this paper, we study the subwavelength imaging (in a conventional sense) properties of the Maxwell's fish eye lens in wave optics. We show that perfect imaging can not be achieved due to the fact that high order modes of the object field will decay quickly before reaching the image point in the Maxwell's fish eye. However, it can still give a subwavelength image and the image resolution is determined by the image field spot size corresponding to the zeroth order component of the object field. Both explicit analysis and FEM simulation are used and they consist very well.

## 2. Mode analysis for a line current in Maxwell's fish eye

Maxwell's fish eye has the following refraction index profile [2]:

$$n = \frac{2n_0}{1+(r/R_0)^2} \tag{1}$$

where $n_0$ and $R_0$ are the refraction index constant and radius of the reference sphere, respectively, and $r$ is the distance between a space point and the center of Maxwell's fish eye. First we set a field source at the center of Maxwell's fish eye. The Helmholtz equation for the field (with a vacuum wave number $k = \dfrac{\omega}{c} = \dfrac{2\pi}{\lambda}$) in a cylindrical coordinate system whose origin is located at the center of Maxwell's fish eye in 2D space can be written as:

$$\frac{1}{r}\frac{\partial}{\partial r}\left(r\frac{\partial E_k(r)}{\partial r}\right) + \frac{1}{r^2}\frac{\partial^2 E_k(r)}{\partial \theta^2} + n^2 k^2 E_k(r) = g(r,\theta) \tag{2}$$

where $g(r,\theta)$ is the source term and $g(r,\theta) = 0$ (when $r \neq 0$). Through variables separation $E_k(r) = E_r(r)E_\theta(\theta)$ and the following variable substitution:

$$\zeta(r) = \frac{r^2 - R_0^2}{r^2 + R_0^2} \tag{3}$$

$$R_0^2 n_0^2 k^2 = v(1+v) \tag{4}$$

equation (2) becomes (we first consider space point $r \neq 0$):

$$\frac{\partial^2 E_\theta}{\partial \theta^2} + m^2 E_\theta = 0 \tag{5}$$

$$(1-\zeta^2)\frac{\partial^2 E_r(r)}{\partial \zeta^2} - 2\zeta\frac{\partial E_r(r)}{\partial \zeta} + \left[v(1+v) - \frac{m^2}{1-\zeta^2}\right] E_r(r) = 0 \tag{6}$$

The solution of Eq. (5) can be expressed as $E_\theta(\theta) = e^{im\theta}$, where $m$ is the mode order of angular momentum. Equation (6) is the general Legendre equation, whose solution is a superposition of the associated Legendre functions. Thus if we set a source at the origin in Maxwell's fish eye medium the general solution of the field is:

$$E_k(r) = \sum_{m=0}^{+\infty}[a_m P_v^m(\zeta(r)) + b_m P_v^m(-\zeta(r))]e^{im\theta} \tag{7}$$

Here we see that the field distribution in Maxwell's fish eye medium can be expressed as a superposition of different order modes $m$. $m \neq 0$ and $m = 0$ represent the high order modes and the zero order mode, respectively, and the high order modes correspond to high angular frequency components. Different sources can excite different modes. If we set some kind of source (at the center of Maxwell's fish eye) that can only excite the zero order mode, the field distribution in Maxwell's fish eye can be written as:

$$E_k(r) = a_0 P_v(\zeta(r)) + b_0 P_v(-\zeta(r)) \tag{8}$$

For a given vacuum wavelength $\lambda$, quadratic equation (4) for $v$ gives two different values of $v$: $v_1$ and $v_2$ (satisfying $v_1 + v_2 = -1$). Since $P_v(\zeta) = P_{-1-v}(\zeta)$, expression (8) for the zero order

mode can be expressed as:

$$E_k(r) = a_0 P_{v_1}(\zeta) + b_0 P_{v_1}(-\zeta) = A E_{v_1}(r) + B E_{v_2}(r) \tag{9}$$

where

$$E_{v_l}(r) = \frac{P_{v_l}(\zeta(r)) - e^{iv_l\pi} P_{v_l}(-\zeta(r))}{4\sin(v_l\pi)} \quad (l=1, 2) \tag{10}$$

and $\dfrac{A}{4\sin(v_1\pi)} + \dfrac{B}{4\sin(v_2\pi)} = a_0$, $\dfrac{e^{iv_1\pi}A}{4\sin(v_1\pi)} + \dfrac{e^{iv_2\pi}B}{4\sin(v_2\pi)} = -b_0$.

Leonhard has shown that a field distribution described by expression (10) has such properties [4]: $E_{v_l}(r) \sim \dfrac{\ln r}{2\pi}, r \to 0$, $E_{v_l}(r) \sim e^{iv_l\pi} \dfrac{\ln r}{2\pi}, r \to \infty$ and $\int_{-\infty}^{+\infty} E_{v_l} e^{-ikt} dk = 0$ for $t < 0$.

Thus, both $E_{v_1}(r)$ and $E_{v_2}(r)$ satisfy the causality and indicate that Maxwell's fish eye can give an image spot at an infinity point if we set a line current at the center. However, they give different field distributions as shown in figure 1. Thus, if we set a line current at the center of Maxwell's fish eye, we can excite zero order mode (8) which can be treated as the superposition of $E_{v_1}(r)$ and $E_{v_2}(r)$. This means a line current at the center of Maxwell's fish eye can generate a zero order mode and a field image can be formed at a point of infinity. The spot size of the field image is determined by the relation between $a_0$ and $b_0$. $a_0$ and $b_0$ should be determined by the boundary condition and the source intensity. Since we have no boundary condition in Maxwell's fish eye, we cannot determine the relation between $a_0$ and $b_0$. This leads to the ambiguity of the image spot size for the zero order mode. The spot size of the image field cannot be infinitely sharp due to the fact that a linear superposition of $P_v(\zeta(r))$ and $P_v(-\zeta(r))$ cannot be a Dirac function at $r \to \infty$. Therefore, the resolution of Maxwell's fish eye for the zero order mode is limited. In section 4 we will discuss the case of high order modes.

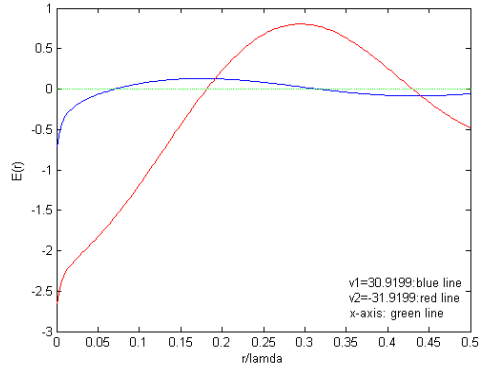

Figure 1. The field distribution (around the source) described by substituting $v_1$ (blue line) and $v_2$ (red line) into expression (10). Here we choose $R_0 = 5\lambda$, $\lambda = 0.2m$ and $n_0 = 1$. From Eq. (4) we obtain

$v_1 = 30.9199$ and $v_2 = -31.9199$.

When the line current is not at the center of the Maxwell's fish eye, we can use the following Möbius transformation to calculate the field distribution [4]:

$$w(z) = -z_\infty \frac{z - z_0}{z - z_\infty} \tag{11}$$

where $z_\infty = -1/z_0^*$, $z_0$ and $z_\infty$ are the positions for the line current and its image. By Möbius transformation, we can obtain the field distribution when the line current is located at $z_0$ in a 2D Maxwell's fish eye:

$$E_k(z) = a_0 P_v(\xi(z)) + b_0 P_v(-\xi(z)) \tag{12}$$

where

$$\xi(z) = \frac{|w(z)|^2 - R_0^2}{|w(z)|^2 + R_0^2} \tag{13}$$

From equations (11) and (12), we can see $a_0$ and $b_0$ are independent of source position $z_0$. However, the zero order mode field produced by a line current is different when the line current is located at different places (i.e., different $z_0$) of Maxwell's fish eye. In general, the zero order mode field around the line current and its image is not of circular shape. The spot size of the zero order mode around the line current and its image are also finite and still related to constants $a_0$ and $b_0$.

### 3. Condition for subwavelength imaging for the object field of zero order mode

In a modified fish eye, we can use PEC boundary condition to determine the relation between $a_0$ and $b_0$, and consequently calculate the finite spot size of the zero order mode around the line current and its image. By using the method of images [4], we can obtain the field distribution for the zero order mode in the modified fish eye when the line current is located at $z_0$:

$$E_k'(z) = E_k(z) - E_k(R/z^*) \tag{14}$$

where $R$ is the radius of the modified fish eye. When we set the line current at the center of the modified fish eye, we have $z_0 = 0$ and $z_\infty = \infty$. By using equations (11), (13), and (14), we can obtain the field at the center of the modified fish eye

$$\lim_{r=|z|\to 0} E_k'(z) = (a_0 - b_0)[P_v(-1) - P_v(1)] \tag{15}$$

On the other hand, we can also determine this value by setting boundary condition $E_k(r) = 0$ at

$r = R$ in equation (8) (without using the method of images). This leads to $a_0 P_v(\frac{R^2 - R_0^2}{R^2 + R_0^2}) + b_0 P_v(-\frac{R^2 - R_0^2}{R^2 + R_0^2}) = 0$. Thus the field produced by a line current at the center of the modified fish eye can be written as:

$$E_k'(z) = a_0 [P_v(\xi) - \frac{P_v(\frac{R^2 - R_0^2}{R^2 + R_0^2})}{P_v(-\frac{R^2 - R_0^2}{R^2 + R_0^2})} P_v(-\xi)] \qquad (16)$$

Consequently, the field at the center of modified fish eye can also be expressed as:

$$\lim_{r=|z|\to 0} E_v'(z) = a_0 [P_v(-1) - \frac{P_v(\frac{R^2 - R_0^2}{R^2 + R_0^2})}{P_v(-\frac{R^2 - R_0^2}{R^2 + R_0^2})} P_v(1)] \qquad (17)$$

If we set a line current at the center of modified fish eye, the field distribution should be the same regardless we use the method of images or not. Thus equations (15) and (17) should be identical, i.e.,

$$b_0 = 0 \qquad (18)$$

$$P_v(\frac{R^2 - R_0^2}{R^2 + R_0^2}) = P_v(-\frac{R^2 - R_0^2}{R^2 + R_0^2}) \qquad (19)$$

When $R = R_0$, condition (19) is satisfied. This means that when the radius of the PEC boundary equates the radius of the reference sphere of Maxwell's fish eye, an image spot can be formed if we set a line current in the modified fish eye. When $R \neq R_0$, a single line current source may give many image spots (as shown later at the end of this section). And the shape of those images may be different from the shape of the object. When equation (19) is satisfied, by substituting (18) and (12) into Eqs. (14), we can obtain the following field distribution in a modified fish eye when the line current is located at $z_0$:

$$E_k'(z) = a_0 [P_v(\xi(z)) - P_v(\xi(R/z^*))] \qquad (20)$$

where $a_0$ is determined by the intensity of the source, and $\xi(z)$ can be determined by Eq. (13).

For example, we choose $R = R_0 = 5\lambda$ and $\lambda = 0.2m$, and set a line current at $z_0(-0.5m, 0)$. We can use Eqs. (11), (13) and (20) to calculate the field distribution in the modified fish eye:

$$E_k(z) = a_0 [P_v(\frac{3r^2 + 8r\cos\theta - 3}{5r^2 + 5}) - P_v(\frac{-3r^2 + 8r\cos\theta + 3}{5r^2 + 5})] \qquad (21)$$

Comparing with Leonhardt's solution [4]:

$$E_k'(z) = \frac{a_0}{4\sin(v\pi)}\{[P_v(\xi(z)) - P_v(\xi(R/z^*))] - e^{i\pi v}[P_v(-\xi(z)) - P_v(\xi(R/z^*))]\}, \quad (22)$$

The results of our analytical solutions (20) and (21) agree well with FEM simulation results as shown in Fig. 2. We can see the spot size around the image point is $FWHM = 0.2925\lambda$ (indicating subwavelength image), which is larger than the spot size around the source point $FWHM = 0.173\lambda$.

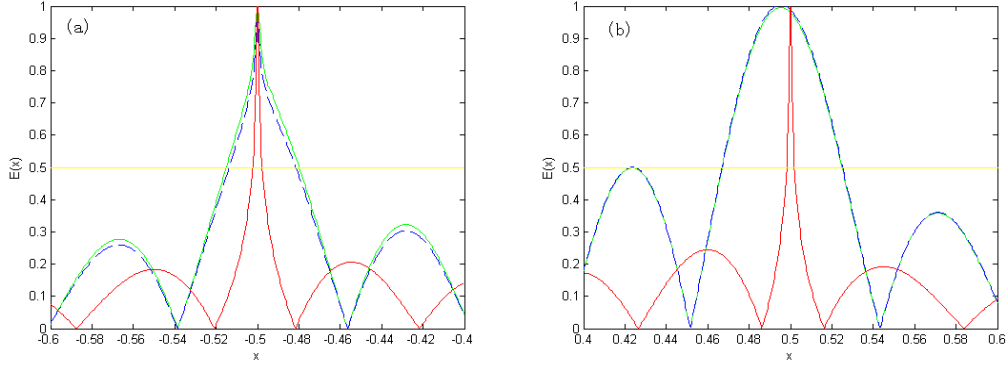

Figure 2 The absolute value of the normalized field distribution around the line current (a) and its image (b) in x direction: green line is from the FEM simulation result when we set a line current at (-0.5m, 0); blue dashed line is our analytical result of Eq. (21); red line is Leonhardt 's analytical result of Eq. (22) for a situation when one sets a line current source at (-0.5m,0) and a line current drain at (0.5m,0). The structure of modified fish eye is $R = R_0 = 5\lambda$ and $n_0 = 1$. The incident wave length is $\lambda = 0.2m$.

From this figure we see that Maxwell's fish eye lens can give a good image of subwavelength resolution if only zero order mode is excited, however, the spot-size of the image field is still larger than the spot size of the source field (indicating that it can not give a perfect image). Adding a special line current drain at the image point [4] can sharpen the image spot size for a very special excitation of object field with only zeroth order mode. However, it is not practical to add active drains in a real imaging application, as we don't know beforehand the distribution of object/source points and consequently cannot determine where to put the active drains. Furthermore, a simple line drain can not produce enough high order modes to make the image as sharp as one wishes (for perfect image) though it can help to recover a very special object field distribution around the image position. We will not discuss the situation of active drains in this paper.

When condition (19) cannot be satisfied, a line current may give many image spots as shown in figure 3. We may also obtain some interesting field pattern that may have some other applications (figure 4)

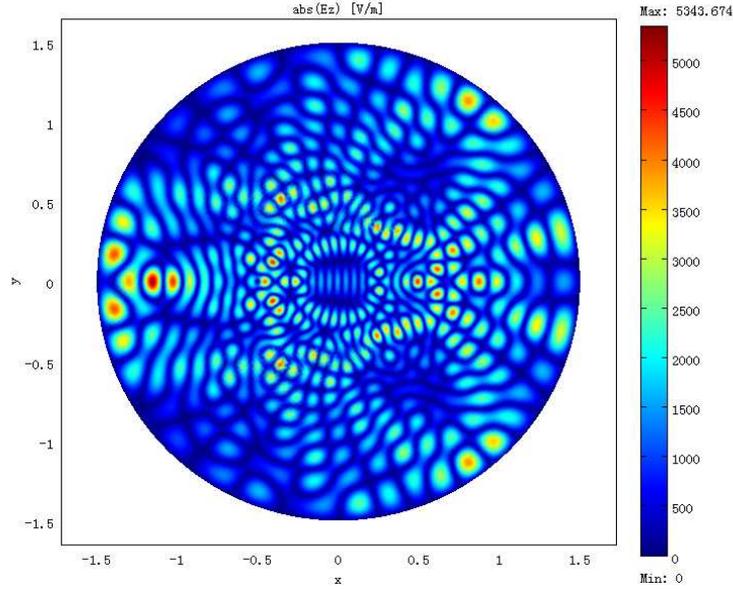

Figure 3 FEM simulation results: the absolute value of the field distribution in the modified fish eye with $R = 7.5\lambda$, $R_0 = 5\lambda$, $\lambda = 0.2m$, $n_0 = 1$ and $z_0 = -0.5m$.

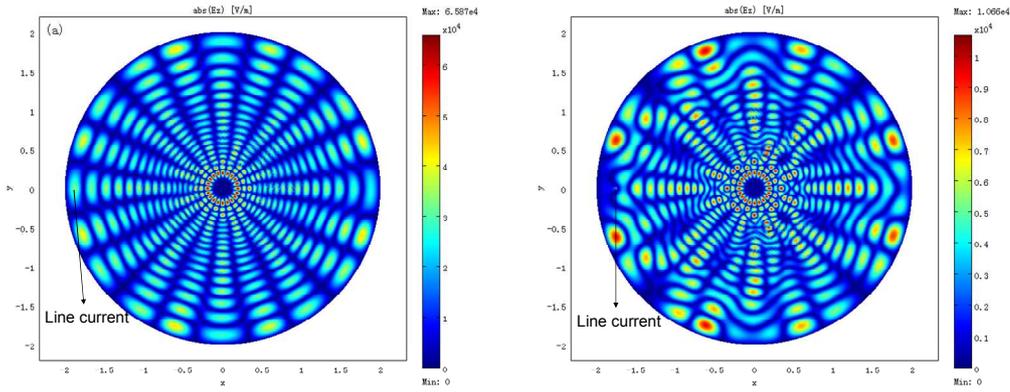

Figure 4 FEM simulation results: the absolute value of the field distribution in the modified fish eye with $R = 10\lambda$, $R_0 = 5\lambda$, $\lambda = 0.2m$, $n_0 = 1$ (a) when we set a line current at (-1.85m,0) (b) when we set a line currents at (-1.75m,0)

## 4. Case for object fields with high order modes

First we show that if we put at the center of original Maxwell's fish eye (without modification) a source (e.g. $\delta(r)f(\theta)$) that can produce high order mode of angular momentum, we cannot get an image spot for those high order modes. The dispersion relationship in a cylindrical coordinate system whose origin is located at the center of Maxwell's fish eye can be written as [6]:

$$k_r^2 + k_\theta^2 = n^2 k^2 \qquad (23)$$

where $k_r$ is the radial component of the wave vector, and $k_\theta$ is the tangential component of the wave vector. Considering the conservation of angular momentum for the m-th order mode:

$$rk_\theta = m \qquad (24)$$

For a high order mode $m \neq 0$ in the Maxwell's fish eye, there is still a caustic [6]. When $m \neq 0$, from equation (24) we can see that $k_\theta$ increases toward the center. Consequently, we can see from Eq. (23) that radial component $k_r$ varies from a real value to an imaginary value as $r \to 0$. The turning point of $k_r = 0$ is the radius of the caustic $R_c$. Inside the caustic, $k_r$ is an imaginary number and the angular momentum state become evanescent (i.e., decay quickly) along the radial direction. The detailed information carried by the high order modes can hardly propagate to the far field without great damping. Only the zero order mode ($m = 0$), which doesn't have the caustic, can propagate to the far field in Maxwell's fish eye. Thus, if we put at the center of Maxwell's fish eye a special source that can excite only (or mainly) high order mode, the field cannot go to the far field, and consequently a subwavelength image can not be formed. If we transform this source position to another point of Maxwell's fish eye or add PEC boundary to Maxwell's fish eye, the situation remains the same: subwavelength image can not be achieved.

We can use FEM simulation to verify this in a modified fish eye with $R = R_0 = 10\lambda$ and $n_0 = 1$. Our simulation is for TE wave in 2D space with $\lambda = 0.2m$. This structure satisfies condition (19). We set a small circle (with radius $r_0$) located at $z_0(-0.5m, 0)$ with boundary condition $E = 3\exp(i\gamma\theta')$ V/m to introduce high order mode. We first choose $r_0 = 10^{-3}\lambda$ and $\gamma = 0$ and the simulation result is shown in Fig. 5. Note that the field generated by boundary condition $E = 3$ on this small circle will contain some high order mode (and thus the object field is quite sharp as compared to Fig. 2(a)), as the zero order mode produced by a line current at (-0.5m, 0) is not a circle as we have discussed in Section 2. Since it also contains some zero order mode, a good image spot can still be formed. However, if we change $\gamma = 0$ to $\gamma = 5$, the situation will be completely different. The simulation result is shown in figure 6. Boundary condition $E = 3\exp(i5\theta')$ on a small circle gives more energy to high order modes (the object field is very sharp). These high order modes cannot propagate to the far field (the ratio of the field around the object to the field around the image position is about $E_o / E_i \sim 10^5$). Consequently, good image can not be achieved, as shown in Fig. 6(b).

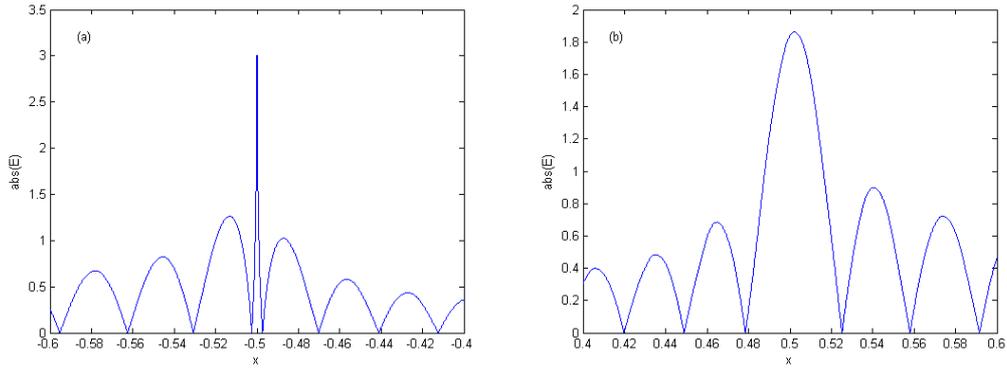

Figure 5. FEM simulation results for the absolute value of the field distribution around the object (a) and its image (b) in x direction for a modified fish eye with $R = R_0 = 5\lambda$, $n_0 = 1$. The field is excited with boundary condition $E = 3\exp(i\gamma\theta')$ V/m at small circle r= $r_0$ with $\gamma = 0$ and $r_0 = 10^{-3}\lambda$. Here we choose $\lambda = 0.2m$.

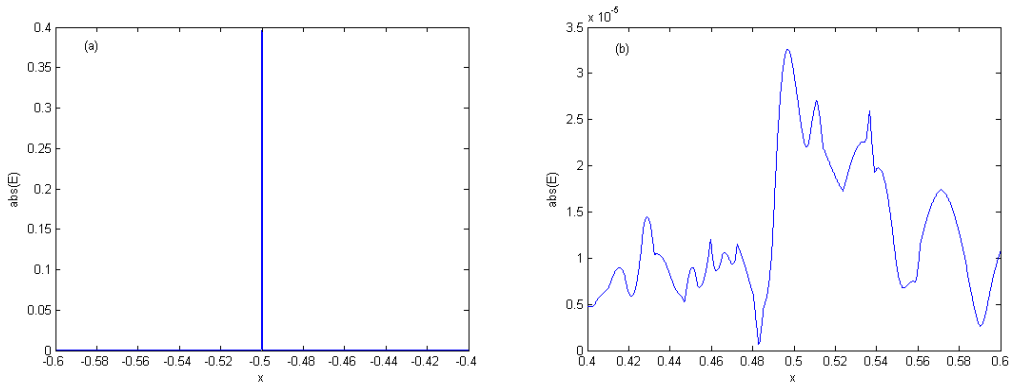

Figure 6. FEM simulation results for the absolute value of the field distribution around the object (a) and its image position (b) in x direction for the same modified fish eye. We have set $\gamma = 5$ (while keeping the other parameters the same as those in Fig. 5) to excite more energy to high order modes $\gamma = 5$

## 6. Summary

Maxwell's fish eye and its modified lens can give a good image of subwavelength resolution (less than 0.3 wavelength) for a line current which only excites zero order mode. However, as we have shown in the present paper that such a Maxwell's fish eye lens cannot give perfect imaging as high order modes of the object field are evanescent modes and can hardly reach the far-field image point. The image resolution is determined by the field spot size of the image corresponding to the zeroth order component of the object field. Both explicit analysis and FEM numerical simulation have been used and they agree very well with each other.

**Acknowledgment**

We thank Prof. Romanov, Dr. Yi Jin, Pu Zhang, Yingran He and Jianwei Tang for helpful discussions.   Email: sailing@kth.se